\documentclass[aps, prl, superscriptaddress,twocolumn,showpacs]{revtex4-1} 
\usepackage{graphicx}
\usepackage{xcolor}
\usepackage{amsmath}
\usepackage{subcaption}
\graphicspath{{Figures/}}

\begin{document}

\title{Thermalization in a Spin-Orbit coupled Bose gas by enhanced spin Coulomb drag}


\author{D.~J.~Brown}\email{dylan.brown@oist.jp}\affiliation{Dodd-Walls Centre for Photonic and Quantum Technologies, Department of Physics, University of Auckland, Private Bag 92019, Auckland, New Zealand}\affiliation{Present address: Light-Matter Interactions for Quantum Technologies Unit,
Okinawa Institute of Science and Technology Graduate University, Onna, Okinawa 904-0495, Japan}
\author{M.~D.~Hoogerland}\affiliation{Dodd-Walls Centre for Photonic and Quantum Technologies, Department of Physics, University of Auckland, Private Bag 92019, Auckland, New Zealand}

\begin{abstract} 
An important component of the structure of the atom, the effects of spin-orbit coupling are present in many sub-fields of physics. Most of these effects are present continuously. We present a detailed study of the dynamics of changing the spin-orbit coupling in an ultra-cold Bose gas, coupling the motion of the atoms to their spin. We find that the spin-orbit coupling greatly increases the damping towards equilibrium. We interpret this damping as spin drag, which is enhanced by spin-orbit coupling rate, scaled by a remarkable factor of $8.9(6)$~s. We also find that spin-orbit coupling lowers the final temperature of the Bose gas after thermalization.
\end{abstract}
\pacs{}
\maketitle

\section{Introduction}
    The understanding of the transport, diffusion and damping of spin, in contrast to those of charge, is important to the field of spintronics \cite{Wolf1488}, where the spin of particles, rather than the quantity of particles (charge) carries information. Spin currents, in contrast to charge currents, are damped due to collisions between particles of opposite spin, as their relative momentum is not conserved. This damping is known as {\em Spin Drag} \cite{PhysRevB.62.4853,Weber:2005vs}.
    
    Analogous to the spin drag in bilayer electron systems, systems of ultracold bosons can also demonstrate spin drag, with the drag enhanced by the familiar bosonic enhancement~\cite{stoof} prominent in ultracold boson systems.
    There is a detailed collection of work over the years into the presence of spin Coulomb drag in ultracold atomic systems ~\cite{Sommer2011, PhysRevLett.111.190402, PhysRevLett.120.170401} with a small selection of the work featuring the inclusion of spin-orbit coupling~\cite{Li2019}.
    
    The other important effect in two dimensional electron systems is spin-orbit coupling (SOC)~\cite{PhysRevB.57.11911,Inarrea:2017vi, chuanzhou2016}. However, solid state materials used to investigate the effects of spin-orbit coupling are often challenging due to the limited control of individual experimental parameters. The body of work surrounding surrounding the topic of spin-drag with added spin-orbit coupling is limited with theoretical investigations looking at the impact of weak coupling on the drag in a 2D electron system~\cite{PhysRevB.75.045333}, or the behaviour of impurities in a spinor condensate system~\cite{PhysRevA.93.023625}.

    In depth understanding of this combination could lead to better understanding of systems such as the topological insulators~\cite{Kane, Bernevig1757} with their famed protected edge states are dependent on the spin-momentum locking caused by the SOC within the material and have been investigated as a potential platform for fault tolerant quantum computation~\cite{Nayak, Raussendorf_2007}.
    
    Ultracold atoms provide an ideal environment for testing the effects of SOC on the spin coulomb drag in quantum systems due to the ability to control many of the crucial parameters accurately. 
    
	Previous experimental and theoretical work by Li et. al.~\cite{Li2019} demonstrated the generation of spin currents using the same technique of a quench of a spin-orbit coupled Bose-Einstein condensate (BEC) and investigated the increased damping of the out of equilibrium system. GPE simulations showed good qualitative agreement with the experimental results, and gathering insight into the BEC shape oscillations and the miscible-imiscible phase transition. As stated by Li et. al. the simulations underestimate the damping of the BEC oscillations, potentially due to the lack of thermal atoms in the simulations. In particular, it was shown in \cite{stoof} that the spin-Coulomb drag between thermal atoms and the condensate dominates over the mean field effects, rendering the mean field GPE only partially effective.

    

    In this article we present our experiments on investigating the thermodynamic behaviour of spin-orbit coupled systems within a conservative potential, and attempt to explain the enhanced damping of the atomic oscillations in the presence of SOC as Coulomb spin drag \cite{stoof} by comparing the results to theoretical calculations.
    
    We create synthetic SOC using the ground state manifold of a Rubidium-87 ($^{87}$Rb) BEC, following the Raman laser scheme first demonstrated in the experiments of Spielman \textit{et. al}~\cite{Lin2011}.
    
	
	A bias magnetic field induces a Zeeman shift, breaking the degeneracy of the F=1 ground state of the atom separating them in energy. 
	A quadratic Zeeman shift $\epsilon$ shifts the $m_F=+1$ state further than the $m_F=-1$, allowing us to effectively decouple the latter from the system. 
	The atoms in the different Zeeman sublevels also differ in momentum by $\delta k_y = 2 k_R$, where $k_R = 2\pi/\lambda$ is the recoil momentum gained by the atom due to absorption of a photon with wavelength $\lambda$. The Hamiltonian of the coupled $F=1$ state as a function of the atomic quasimomentum $\tilde{k}_y$ , is as follows, 

        \begin{equation}\label{eq:soc3levelH}
        	\hat{H}_y(\tilde{k}_y)=
        	\begin{pmatrix}
        	\frac{\hbar^2(\tilde{k}_y+2k_R)^2}{2m}-\hbar\delta & \frac{\hbar\Omega_R}{2} & 0\\
        	\frac{\hbar\Omega_R}{2} & \frac{\hbar^2\tilde{k}_y^2}{2m}-\hbar\epsilon & \frac{\hbar\Omega_R}{2}\\
        	0 & \frac{\hbar\Omega_R}{2} & \frac{\hbar^2(\tilde{k}_y-2k_R)^2}{2m}+\hbar\delta
        	\end{pmatrix}
    \end{equation}

    Diagonalizing the Hamiltonian gives the energies of the spin-orbit coupled dressed states which for Raman coupling strengths $\hbar\Omega_R < 4E_R$ features a double minimum. Here, $E_R = \hbar^2 k_R^2 /2m$, $\delta$ is the two photon detuning  between the bare states, and $\epsilon$ indicates the quadratic shift. Correctly choosing the detuning for a given coupling results in the ground state being an equal superposition of the two pseudospin states $|\uparrow\rangle$, $|\downarrow\rangle$, corresponding to a spin-orbit coupled state. 
    
    However, when $\hbar \Omega_R > 4 E_R$, there is only a single minimum at quasimomentum $\tilde{k}_y=k_R$. In the experiments reported here, we initially prepare a Bose-Einstein Condensate, trapped in a harmonic trap, in the latter state with $\hbar \Omega_R = 5.5 E_R$. The system is then quenched to a lower $\hbar \Omega_R < 2E_R$, which takes the system out of equilibrium, and allowed to thermalise. We find the time constant for this thermalisation, and find that the rate scales with the coupling strength $\Omega_R$. 

\section{Experimental apparatus}
    \begin{figure}[t]
    \includegraphics[width=\columnwidth]{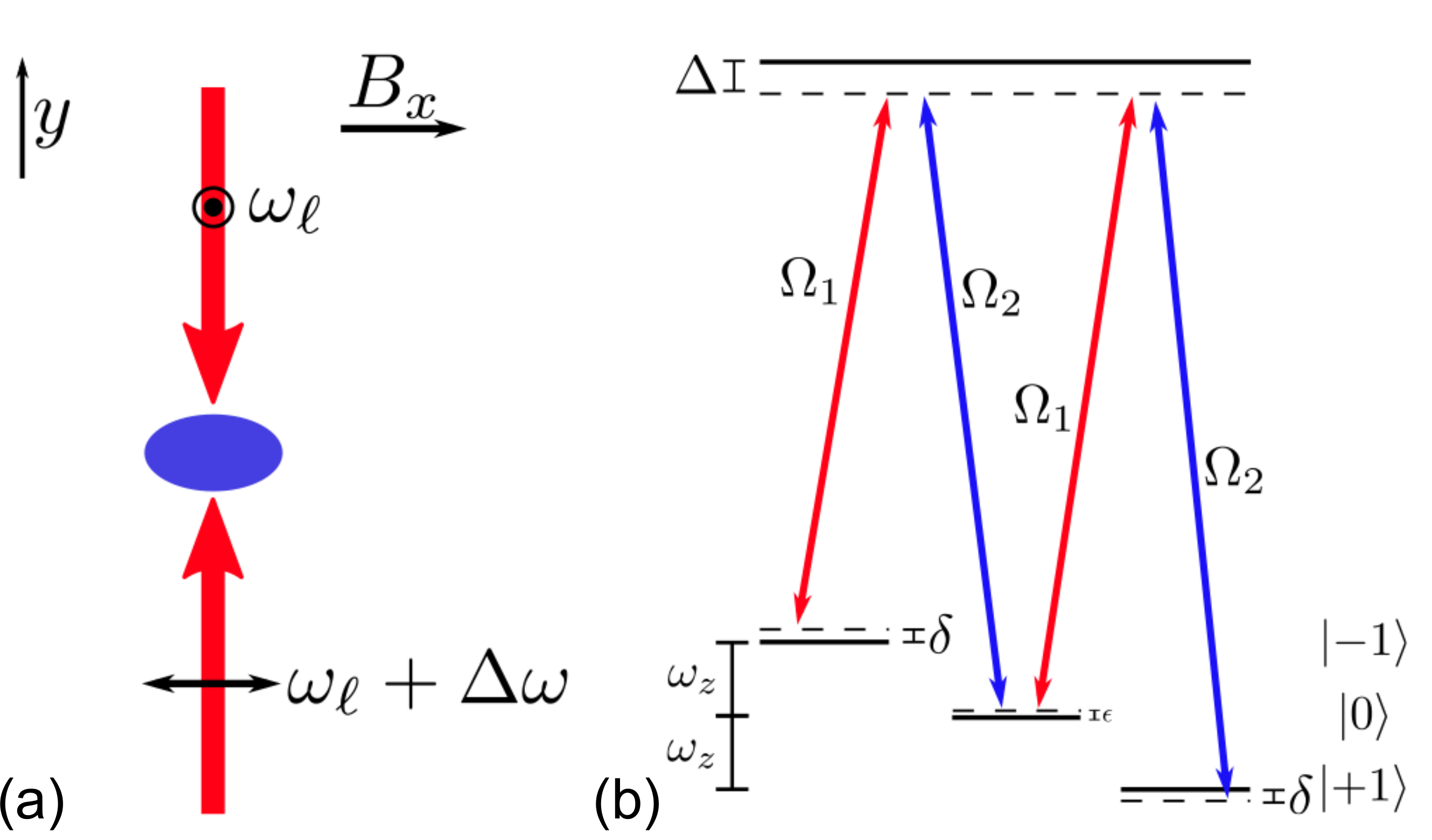}
    \caption{Schematic of the experimental setup. (a) Geometry of the laser orientation and polarization, along with the bias magnetic field.
    (b) The level scheme of the $F=1$ ground state and the Raman transitions induced by the coupling lasers. The Rabi frequencies of the individual transitions are $\Omega_1$ and $\Omega_2$.}  \label{fig:expt}
    \end{figure}
    
    Our experiments begin with an all-optical BEC composed of approximately 10$^4$ $^{87}$Rb atoms, optically pumped into the $|F=1;m_F=0\rangle$ before evaporation, as described in our previous work~\cite{Brown2018}. The BEC is held in a harmonic trap, with aspect ratios $\omega_y:\omega_x:\omega_z = 1:1.2:2$, formed by a crossed-beam optical dipole trap. We use two values of $\omega_y$ for our experiments. The lowest trapping frequencies correspond to the the experiments performed with the trap held at the final power 66~mW achieved after evaporation, with $\omega_y = 2\pi\times85$~s$^{-1}$. For the larger trapping frequencies, we adiabatically increase the power of the dipole trapping laser to 90~mW, corresponding to $\omega_y=2\pi\times112$~s$^{-1}$. Note that the larger trap frequency corresponds to a larger trap depth, and thus increases the number of atoms retained in the trap during the thermalization process and increases the rate of collisions between the atoms.
    
    During the evaporation to BEC a magnetic bias field B$_y$ is ramped up in the last 2 seconds to 8.35~G providing the a measured $\omega_B/2\pi = 5.845$~MHz Zeeman shift, and a measured quadratic Zeeman shift of $\epsilon/2\pi = 5$~kHz. A schematic of the coupling scheme and geometry is shown in Fig.~\ref{fig:expt}. The two-photon coupling strength $\Omega_R$ is experimentally determined by observing the Rabi oscillations between the populations of the states $|-1\rangle$, $|0\rangle$ and $|+1\rangle$ zero detuning, and fitting the time evolution with the three state optical Bloch equations. 
    
    To induce spin-orbit coupling we use two orthogonally polarized laser beams with wavelength $\lambda=790.2~$nm counter-propagating along $y$ that are focused to an 150~$\mu$m diameter beam onto the center of the dipole trap. This wavelength of $790.2$~nm was chosen to minimise the scalar AC Stark shift in the atoms, which would have led to undesirable extraneous forces induced by these beams. The two beams are derived from the same laser, but differ in frequency by $\Delta\omega\approx\omega_B+\epsilon$ and couple two of the internal $m_F$ levels of the BEC atoms. For sufficient quadratic Zeeman shift, that is $h\epsilon > E_R$ the $m_F=+1$ internal state is tuned out of resonance for the two photon Raman coupling. The coupled system becomes an effective two level system of spin-momentum states which we label $|m_F=-1,\tilde{k}_y+2k_R\rangle = |\uparrow'\rangle$ and $|m_F=0,\tilde{k}_y\rangle = |\downarrow'\rangle$.
    
\section{Experimental Procedure}

\begin{figure}[t]
        \includegraphics[width=\columnwidth]{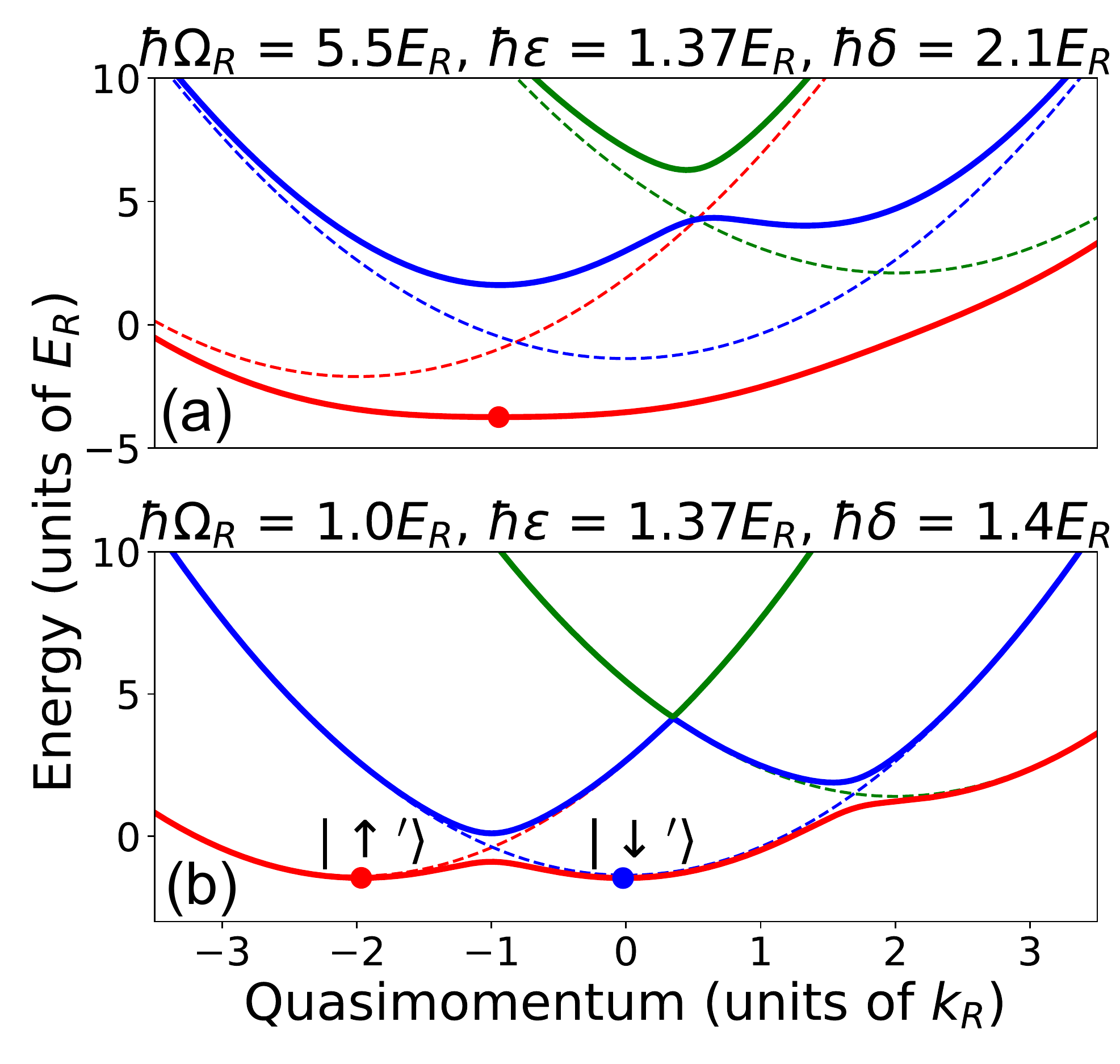}  
        \caption{Dispersion relations for the coupled BEC for two coupling strengths. (a) $\hbar\Omega_R = 5.5E_R$, features a single minima where both spins have equal populations and the same quasimomentum. (b) $\hbar\Omega_R=1E_R$ features two minima of the dispersion. Each pseudospin occupies one of the minima with the corresponding quasimomentum.}
        \label{fig:dispersion}
    \end{figure}

    The condensate is prepared in the lowest energy dressed band of the Raman coupled system by adiabatically increasing the Raman coupling to 5.5~$E_R$ over 50~ms, where there is a single minimum in the dispersion curve as illustrated in Fig.\ref{fig:dispersion}(a). The adiabatic increase of the coupling prevents unwanted heating and oscillations of the condensate in the trap caused by synthetic electric fields~\cite{Lin2011}. We hold the Raman coupling on for a further 30~ms at a constant value in order to ensure the system is in the lowest energy dressed band. We confirm that the ramp is adiabatic by measuring the total momentum of the atoms, obtained from the weighted sum of the quasimomentum of all momentum components, during this 30~ms period and confirming that it is zero at each point in time. If the total momentum is non-zero during this phase, the ramp speed must be adjusted to ensure the atoms remain in the lowest energy dressed state.
    
    To take the system out of equilibrium, a synthetic electric force is imparted on the dressed BEC by abruptly reducing the Raman coupling strength from the initial $\Omega_i=5.5E_R/\hbar$ to a final value $\Omega_f$ in 1~ms. The rapid decrease of Raman coupling constitutes a quench of the system. The condensate separates into the two pseudospin states $|\uparrow'\rangle$ and $|\downarrow'\rangle$ through the synthetic electric force, each accelerating towards one of the new minima (see Fig.~\ref{fig:dispersion}(b)) of the dispersion relation, where they then oscillate in the harmonic trap with maximal momentum $|k_{\uparrow,\downarrow}| = \pm1\hbar k_R$.

	To compensate for both the impact of the $m_F=+1$ state and the changing AC stark shift as the laser intensity changes, we adjust the laser frequency difference to maintain equal populations of the two states, by an amount up to $\hbar\delta_\mathrm{AC}=1E_R$. This shift in the two photon resonance condition is extremely sensitive to small changes in experimental parameters, such as the magnetic fields.  Although care is taken to maintain equal populations of the spin components, the final spin-orbit coupled state after the quench will occasionally have non-equal populations in each component. To group the data we calculate the population imbalance     
	
    	\begin{equation}
    		F_{\uparrow\downarrow} = \frac{N_\uparrow - N_\downarrow}{N_\uparrow + N_\downarrow},
    	\end{equation}
	where $N_\uparrow$ is the population of the $m_F=-1$ state and $N_\downarrow$ is the population of the $m_F=0$ state. 
	In this paper we focus on the case with balanced populations, where $|F_{\uparrow\downarrow}|<0.1$, by post-selecting the data. 
	
    We let the two pseudospin states oscillate in the dipole trap for time $t$ up to 20~ms before switching the trap and Raman coupling off simultaneously, projecting the atoms onto their bare spin-momentum states. The bare states expand for 15~ms in a Stern-Gerlach gradient separating the spin components in the $x$ dimension before being imaged with a resonant absorption method. 
	
	We measure the rate of thermalization of the system by evaluating the momentum distribution of both spin states as a function of time. We numerically determine the mean momentum of each of the spin ensembles as they oscillate in the trap with a decaying amplitude. We fit the decay of the oscillation of the momentum difference $k_t = |k_\uparrow - k_\downarrow|$ and measure the final temperature of the thermalized ensembles.

\section{Thermalization of a Spin-Orbit coupled BEC}

		\begin{figure}[t]
			\includegraphics[width=\columnwidth]{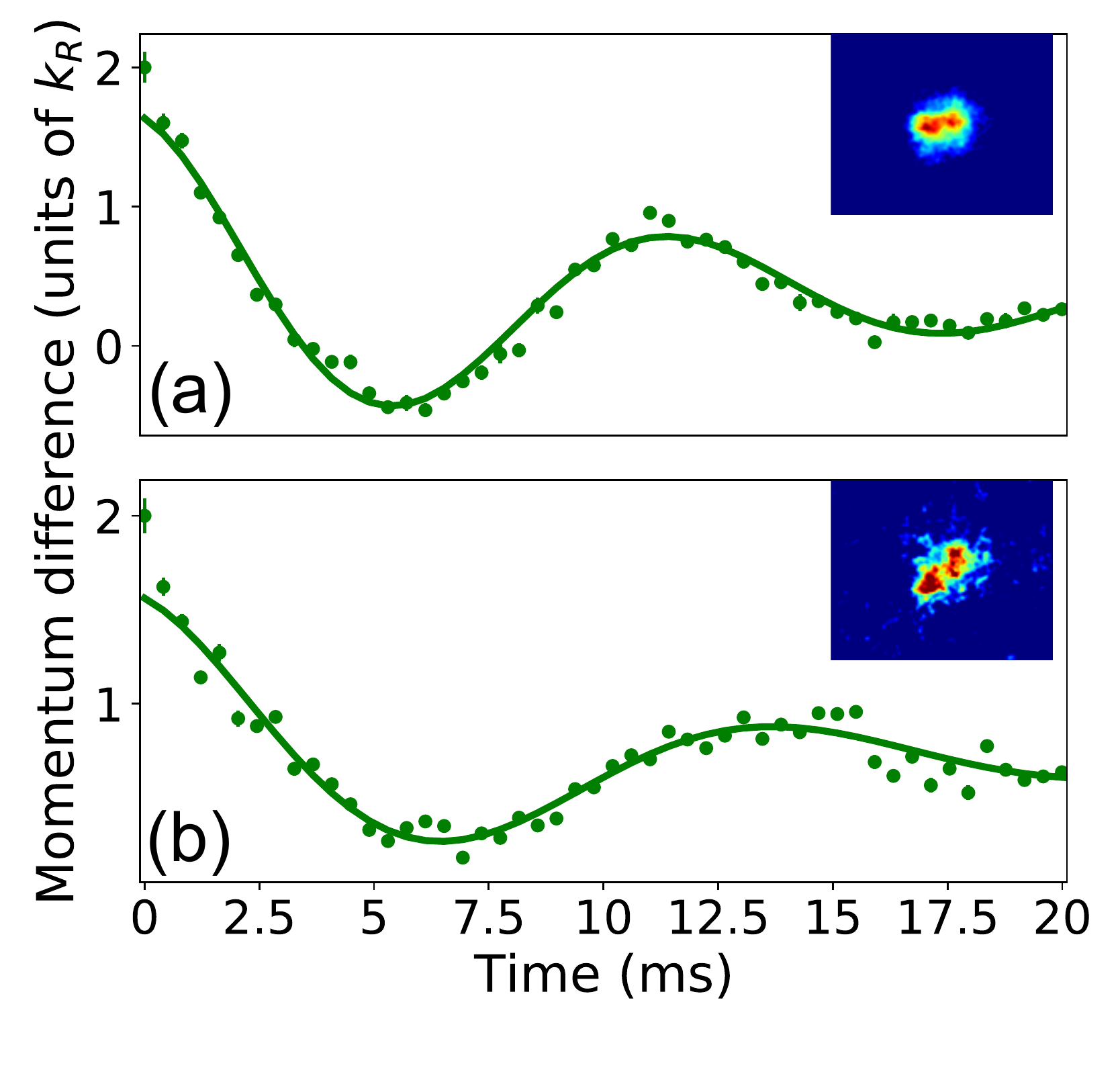}
		\caption{ 
        Plots of the momentum difference between the two components of the system for two coupling strengths (a) $\hbar\Omega_f=1.5E_R$ and (b) $\hbar \Omega_f=2E_R$ fit with a decaying cosine function. The insets show the spin-momentum distribution after the system has returned to equilibrium, and clearly demonstrate the return of the system to a spin orbit coupled state with the final momentum distribution reflecting the non-zero quasimomentum before release from the trap.} 
		\label{fig:zeroPop20} 
	\end{figure}
	
	It is important to note that the momentum imparted on the pseudospins during the quench depends of the final coupling strength, which arises from the quasimomentum minima shifting as a function of the coupling strength. The shift in this work on the order of $\pm0.05\hbar k$ for each spin, accounting for a 5\% difference in the total momentum of the atoms of $\hbar\Omega_f=0$ and $\hbar\Omega_f=2E_R$
    
    Once the system has thermalized and the oscillations have completely damped, a small fraction of the condensate remains, with the atoms occupying the minima of the new spin-orbit coupled band. For Raman coupling above $\hbar\Omega_R=1E_R$ the time-of-flight images show clearly the system has returned to equilibrium in a spin-orbit coupled state with the pseudospin momentum clearly being non-zero. We confirm the non-zero momentum of the atoms comes from the quasimomentum of the spin-orbit coupled state, rather than residual oscillation energy, by noting the momentum remains unchanged over 5~ms of evolution.
    
    Fig.~\ref{fig:zeroPop20} demonstrates two cases where the final pseudospins are separated from zero momentum when reaching equilibrium. A clear example is shown in the inset of Fig.~\ref{fig:zeroPop20}(b) shows the pseudospins are positioned $k_{\uparrow,\downarrow} = \pm0.25\hbar k_R$ , corresponding to quasimomentum before release $\tilde{k}_y = \mp0.75\hbar k_R$, the locations of the dispersion minima obtained from exact diagonalization of the Hamiltonian. Even though the trap frequencies are the same for the two situations in the figure, the dispersion relation is different for different coupling, giving rise to the observed difference in oscillation frequency. It is also clear that the higher coupling strength (b) gives rise to a stronger damping of the oscillation.
    
    As mentioned, some data was also obtained for imbalanced populations. In this case, qualitatively we observe that the smaller population oscillation damps more rapidly while the larger population continues to oscillate. Accurately controlling the imbalanced populations proves to be difficult and therefore we do not include these results in this paper.

\section{Spin Coulomb drag}\label{sec:coulomb}
	
	For a situation with no spin-orbit coupling, the damping coefficient can be determined theoretically for an ultracold Bose gas. The spin drag between two components can be calculated from two expressions for the non-condensed and the condensed atoms respectively \cite{stoof},
    
    \begin{align}
        \nonumber&\gamma_{22} = \frac{n\Lambda a^2_{\uparrow\downarrow}}{\hbar\beta}\frac{1}{6\pi^2}\frac{1}{\left(n\Lambda^3\right)^2} \int_{0}^{\infty}  \frac{\,dq\,d\omega \,q^2}{\sinh^2\left(\omega/2\right)} \\ &\times \ln\left(\frac{\exp\left[q^2/16\pi + \beta gn_0 - \omega/2 + \pi\omega^2/q^2\right] - \exp\left[-\omega\right]}{\exp\left[q^2/16\pi + \beta gn_0 - \omega/2 + \pi\omega^2/q^2\right] - 1}\right),
    \end{align}
    
    and 
    
    \begin{align}
           \nonumber&\gamma_{12} = \frac{n\Lambda a^2_{\uparrow\downarrow}}{\hbar\beta}\frac{64n_0a}{3(2\pi)^3n\Lambda} \int_{0}^{\infty} \,dp_1\,dp_3\,p_1p_3^3 \\ \nonumber&\times\left[1 + \frac{1}{\exp\left[\left(p_1^2 + p_3^2\right)/4\pi +2\beta gn_0\right] - 1} \right] \\ \nonumber&\times \frac{1}{\exp\left[p_1^2/4\pi +\beta gn_0\right] - 1}\times \frac{1}{\exp\left[p_3^2/4\pi +\beta gn_0\right] - 1} \\ &\times\Theta\left(\frac{p_1p_3}{2\pi} - \beta gn_0\right).
    \end{align}
    
    Here, $\Theta$ is the Heaviside function, $\Lambda = \sqrt{2\pi\hbar^2/mk_BT}$ is the thermal de Broglie wavelength, $\beta = 1/k_BT$ the inverse thermal energy and $g = 4\pi\hbar^2a/2m$ the interparticle interaction strength.
    Calculations were performed using a script provided by Jogundas Armaitis~\cite{stoof} with our experimental parameters, returning the total spin drag relaxation rate for a given density of atoms. Due to the fact the atoms are oscillating in the trap and only overlap and only interact periodically, we multiply by a scaling factor calculated based on the interaction time of the two spin clouds overlapping in the trap. 
    
	\begin{figure}[t]
		\centering
		\begin{subfigure}{0.98\columnwidth}	
		       \includegraphics[width=\textwidth]{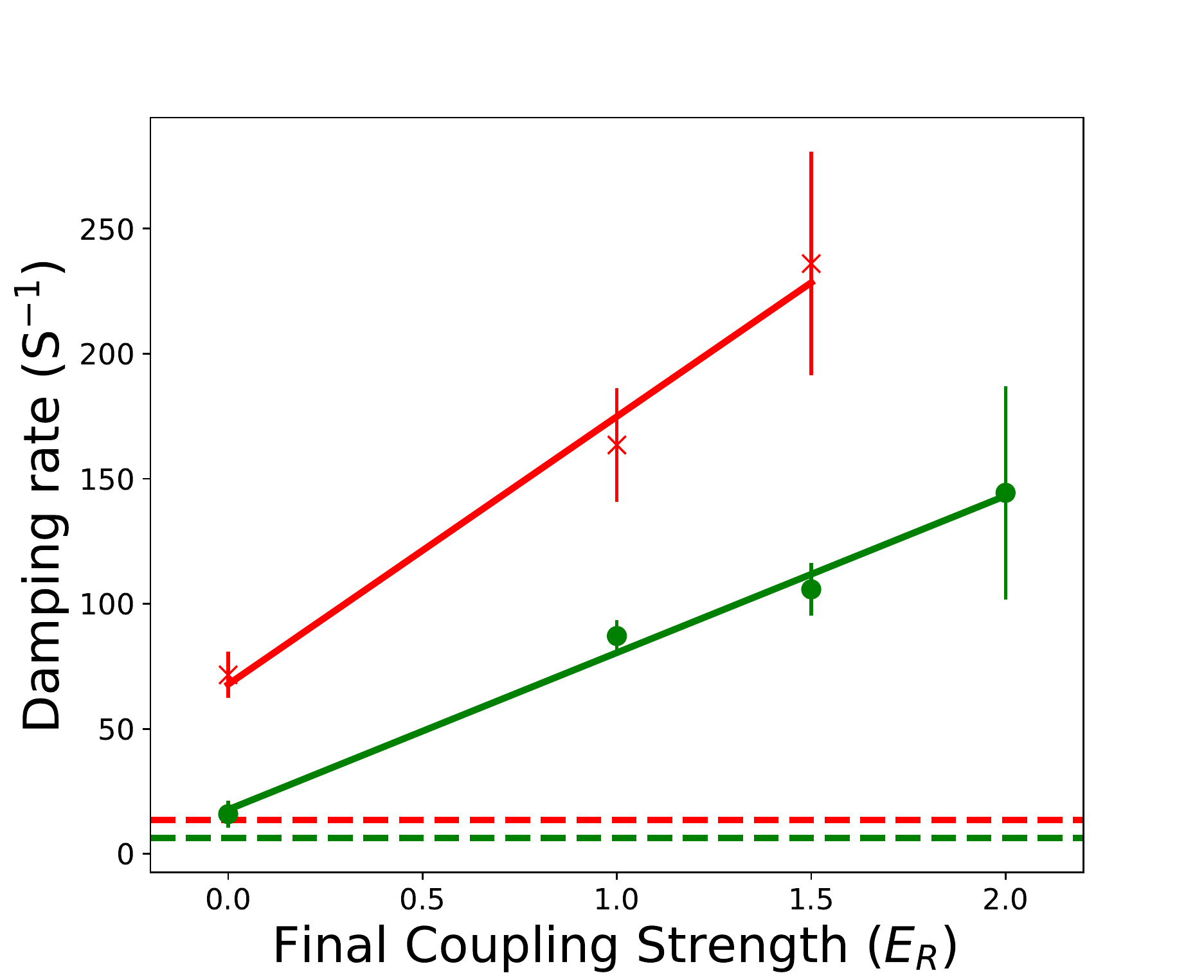}
		\end{subfigure}
		\caption{The damping rate of the system as a function of the Raman coupling strength, for low (green circles) and high (red crosses) trapping frequencies. The damping increases linearly over the range of coupling, with the gradient being a combination of both collisional damping and spin drag.
		The red and green dashed lines indicates the theoretical spin drag damping rates.}
		\label{fig:damping}
	\end{figure}
	
    At the time of writing we are unable to obtain theoretical calculations for the effects of the spin-orbit coupling on the spin drag, so we compare the experiments for the uncoupled case with the theory.
    For the $y$ trapping frequencies $\omega_y = 2\pi\times85$~s$^{-1}$ and $\omega_y = 2\pi\times112$~s$^{-1}$, we calculate a spin drag damping rate of $\gamma_s=\gamma_{12}+\gamma_{22}=$13.6~s$^{-1}$. Comparing this to our experimentally observed damping rate of $\gamma=72(9)$~s$^{-1}$, we observe that part of $\gamma_e$ is caused by collisions of the condensate with thermal atoms, and is dependent on the atomic density and is also present regardless of spin drag.
    
    \begin{figure}[t]
		\centering
			\includegraphics[width=\columnwidth]{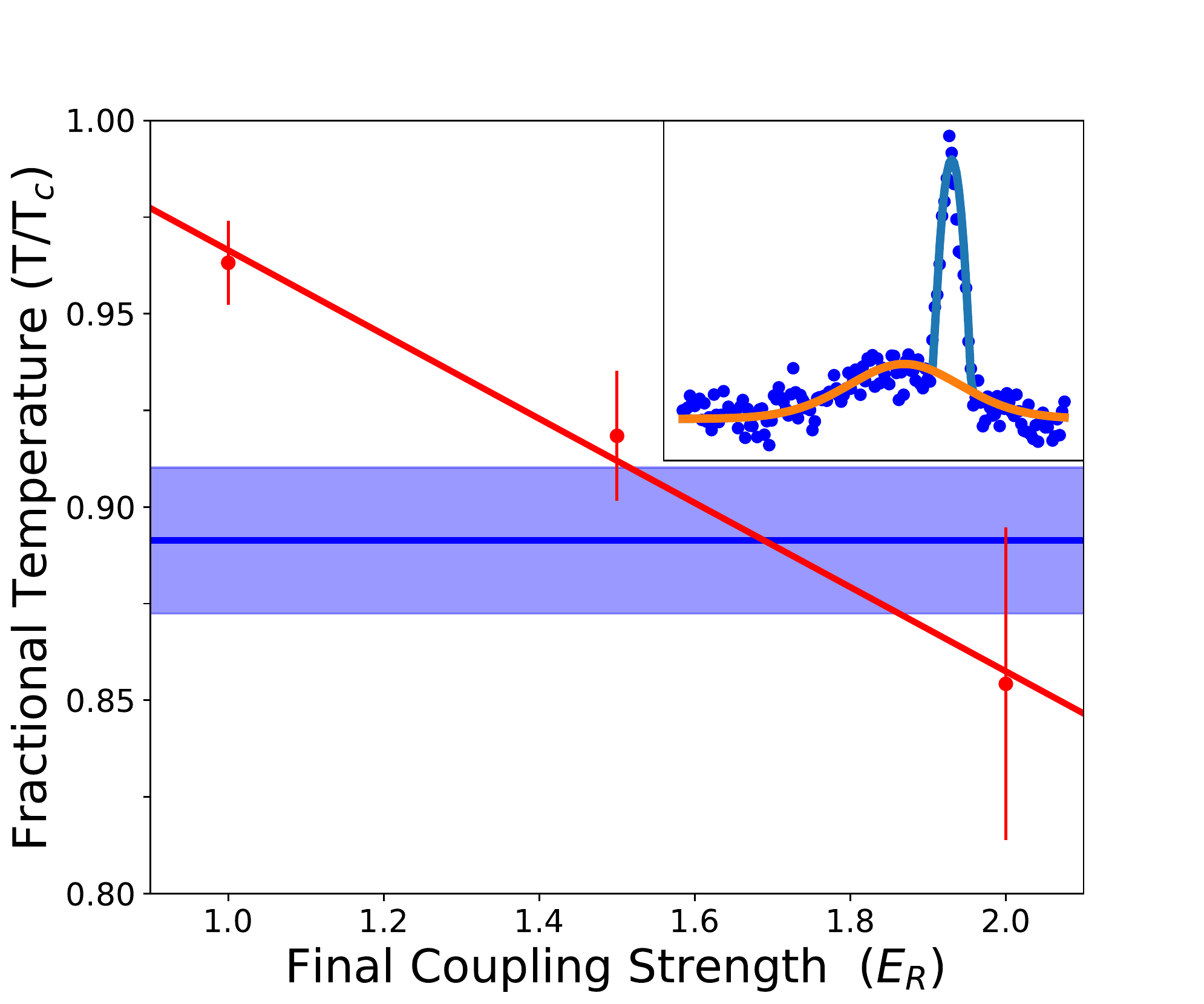}
		
		\caption{Fractional temperature T/T$_c$ of the system as a function of the final coupling strength. The blue line shows the temperature before the quench averaged over many shots, with the shaded region indicating the error. The temperature of the equilibrium state is shown with red circles, demonstrating a clear decrease in the heating for larger coupling strengths. Inset: The integrated 1D profile showing the bimodal fit to one of the spin components}
		\label{fig:temperature}
	\end{figure}
	
	Taking into account the elastic collision rate, it is clear from Fig.~\ref{fig:damping} that the increase in the spin-orbit coupling corresponds to a significantly increased damping rate, with a linear dependence over the range measured. We find that the damping rate also scales with the calculated spin-drag damping rate and e summarize our results for the damping coefficient $\gamma$ by the expression
    \begin{equation}
        \gamma=\gamma_e + \xi\Omega_R\gamma_s
    \end{equation}
    where $\xi$ is a constant. 
    The fit parameters in Fig.~\ref{fig:damping}, as well as the spin drag constant without spin-orbit coupling are summarised in table~\ref{tab:one}. 
    \begin{table}[h]
        \centering
        \begin{tabular}{|c|c|c|c|}
        \hline
            $\omega_y/2\pi$ (s$^{-1}$) & $\gamma_e$ (s$^{-1}$) & $\xi\gamma_s$ & $\gamma_s$ (s$^{-1}$) \\
             \hline
             85 & 18(6) & 63(4) &  6.4\\
             112 &  67(14) & 107(13) & 13.6 \\
             \hline
        \end{tabular}
        \caption{Scaling of the fit parameters in figure~\ref{fig:damping}}
        \label{tab:one}
    \end{table}
    
    At the time of writing, we have not found a way to find $\xi$ from theoretical considerations for our experimental configuration, but from table~\ref{tab:one} we find that $\xi=8.9(6)$~s, which is remarkably large as spin-orbit coupling affects the condensate fraction much more strongly than the thermal fraction. We envision that finding an accurate theoretical value may take a truncated Wigner type simulation \cite{Blakie, Brown2018} to include both spin components, the spin-orbit coupling along with their interactions with both the opposing spin condensate atoms, but also the atoms belonging to the thermal cloud.
    
	Finally, we measure the final temperature of the system once it has reached equilibrium. We integrate the time-of-flight region for each spin to obtain a 1D density profile and fit them with a sum of a Bose enhanced Gaussian and a Thomas-Fermi profile. Integrating the fits we obtain the atom number for the BEC and thermal component, which we use to obtain the fractional temperature T/T$_c$. We plot the measured temperatures in Fig.~\ref{fig:temperature} along with the initial temperature and uncertainty. We fit a straight line to the temperature, obtaining $T/T_c = 1.08(0.01) - 0.11(0.02)\Omega_R$.
	
	It is interesting to note that the condensate is off center with respect to the thermal cloud, consistent with the quasimomentum of the final spin orbit coupled state mentioned in an earlier section.
	The results surprisingly show that for increasing Raman coupling, the damping results in a reduced final temperature, possibly indicating the spin-orbit coupling plays a significant role in the relaxation process. The means by which the temperature decreases is not so obvious, however the condensate fraction remaining at the end of the experiment is increased for increasing coupling. 
	
\section{Conclusion}

    We have presented experiments performed to investigate the impact of spin-orbit coupling on the thermalization processes present in an out of equilibrium system of ultracold bosons. We measure the spin drag damping rate of the atoms and compare the uncoupled case to theoretical calculations. We show that introducing the spin-orbit coupling into the system strongly increases the rate at which the system returns to equilibrium, while also reducing the temperature increase caused by the excitation. Finally, we have shown that the equilibrium state of the system after rethermalization is a spin-orbit coupled BEC, with the quasimomentum measurements after reaching equilibrium corresponding to the dispersion relation calculated through exact diagonalization of the Hamiltonian. We anticipate that this work will lead to new understanding of thermalization in the presence of spin-orbit coupling.

\bibliography{bibliography.bib}

\end{document}